\documentclass{moriond}

\usepackage[numbers,square,super]{natbib}
\usepackage{aas_journals}

\newcommand{\corr}[2]{{\color{magenta}#2}}

\newcommand{\be}{\begin{equation}}
\newcommand{\ee}{\end{equation}}
\newcommand{\ba}{\begin{eqnarray}}
\newcommand{\ea}{\end{eqnarray}}
\newcommand{\tr}{\tilde{r}}

\newcommand{\mpc}{\rm {h^{-1}Mpc }}
\newcommand{\ltsima}{$\; \buildrel < \over \sim \;$}
\newcommand{\lsim}{\lower.5ex\hbox{\ltsima}}
\newcommand{\gtsima}{$\; \buildrel > \over \sim \;$}
\newcommand{\gsim}{\lower.5ex\hbox{\gtsima}}

\def\gtrsim{\mathrel{\hbox{\rlap{\hbox{\lower4pt\hbox{$\sim$}}}\hbox{$>$}}}}

\def\lesssim{\mathrel{\hbox{\rlap{\hbox{\lower4pt\hbox{$\sim$}}}\hbox{$<$}}}}

\newcommand{\LCDM}{$\Lambda$CDM}

\def\be{\begin{equation}}
\def\ee{\end{equation}}
\def\bea{\begin{eqnarray}}
\def\eea{\end{eqnarray}}

\begin{document}
\title{The Cold Spot in the Cosmic Microwave Background: the Shadow of a Supervoid}
\author{Istv\'an Szapudi$^{1}$, Andr\'as Kov\'acs$^{2,3}$, Benjamin R. Granett$^{4}$, Zsolt Frei$^{2,3}$, Joseph Silk$^{5}$, J. Garc\'ia-Bellido $^{6}$, Will Burgett$^{1}$,
Shaun Cole$^{7}$, Peter W. Draper$^{7}$, Daniel J. Farrow$^{7}$, Nicholas Kaiser$^{1}$, Eugene A. Magnier$^{1}$, Nigel Metcalfe$^{7}$,
  Jeffrey S. Morgan$^{1}$, Paul Price$^{8}$, John Tonry$^{1}$, Richard Wainscoat$^{1}$\\
$^{1}$ Institute for Astronomy, University of Hawaii 2680 Woodlawn Drive, Honolulu, HI, 96822, USA\\
$^{2}$ Institute of Physics, E\"otv\"os Lor\'and University, 1117 P\'azm\'any P\'eter s\'et\'any 1/A Budapest, Hungary\\
$^{3}$ MTA-ELTE EIRSA "Lend\"ulet" Astrophysics Research Group, 1117 P\'azm\'any P\'eter s\'et\'any 1/A Budapest, Hungary\\
$^{4}$INAF OA Brera, Via E. Bianchi 46, Merate, Italy\\
$^{5}$Department of Physics and Astronomy, The Johns Hopkins University, Baltimore MD 21218, USA\\
$^{6}$ Instituto de F\'isica Te\'orica IFT-UAM/CSIC, Universidad Aut\'onoma de Madrid, Cantoblanco 28049 Madrid, Spain\\
$^{7}$Department of Physics, Durham University, South Road, Durham DH1 3LE, UK\\
$^{8}$Department of Astrophysical Sciences, Princeton University, Princeton, NJ 08544
}

\maketitle\abstracts{
\bigskip
Standard inflationary hot big bang cosmology predicts small fluctuations in the Cosmic  Microwave Background (CMB) with isotropic Gaussian statistics.  All measurements support the standard theory, except for a few anomalies discovered in the Wilkinson Microwave Anisotropy Probe maps and confirmed recently by the  Planck satellite. The Cold Spot is one of the most significant of such anomalies, and the leading explanation of it posits a large void that imprints this extremely cold area via the linear Integrated Sachs-Wolfe (ISW) effect due to the decay of gravitational potentials over cosmic time, or via the Rees-Sciama (RS) effect due to late-time non-linear evolution. Despite several observational campaigns targeting the Cold Spot region, to date no suitably large void was found at higher redshifts $z > 0.3$. Here we report the detection of an $R =(192 \pm 15)\mpc$ size supervoid of depth $\delta = -0.13 \pm 0.03$, and centred at redshift $z = 0.22$. This supervoid, possibly the largest ever found, is large enough to significantly affect the CMB via the non-linear RS effect, as shown in our Lemaitre-Tolman-Bondi framework. This discovery presents the first plausible explanation for any of the physical CMB anomalies, and raises the possibility that local large-scale structure could be responsible for other anomalies as well.
}

\section{Introduction}

The Cosmic Microwave Background (CMB) Cold Spot (CS) is an exceptionally cold area centered on on $(l,b) \simeq (209^\circ,-57^\circ)$ and was first detected in the Wilkinson Anisotropy Probe \cite{bennett2012} at $\simeq 3\sigma$  significance using wavelet filtering \citep{VielvaEtal2003, CruzEtal2004}. The CS is perhaps the most significant among the ÒanomaliesÓ recently confirmed by {Planck \citep{Planck23}. Given that these potential departures from Gaussianity and/or isotropy have relatively low significance, any of them could represent somewhat rare statistical fluctuations. Nevertheless, they are significant enough to warrant further studies, and indeed they have already spawned a score of interesting ideas. Some of these appeal to hitherto undiscovered physics, e.g., textures \citep{CruzEtal2008,Vielva2010}, while others \citep{InoueSilk2006,InoueSilk2007,InoueEtal2010,Inoue2012} posit that the CS, and possibly other anomalies, are caused by the Integrated Sachs-Wolfe effect (ISW) \citep{SachsWolfe,ReesSciama} of the decaying gravitational potentials, which in turn is caused by Dark Energy. The latter explanation would require of a large, $\gtrsim 200\mpc$ supervoid centered on the CS \citep{InoueSilk2006,InoueSilk2007}, readily detectable in large scale structure surveys.

In the past, there have been several attempts to search for voids in large scale structure data in the direction of the CS. Rudnick et al. \cite{RudnickEtal2007} detected a low density region approximately aligned with the CS, although the detection significance has been disputed \cite{SmithHuterer2010}. Bremer et al.\cite{BremerEtal2010} conducted a redshift survey in the area, and found no evidence for a void in the redshift range of $0.35 < z < 1$, while Granett et al.\cite{GranettEtal2010} conducted an imaging survey with photometric redshifts, and excluded the presence of a large underdensity of  $\delta\simeq -0.3$ between redshifts of $0.5<z<0.9$. Both of these surveys ran out of volume at low redshifts due to their small area, although the  data\cite{GranettEtal2010}are consistent with the presence of a void at $z < 0.3$ with low significance and no void was found at $0.3<z<0.5$, even if the existence of one could not be excluded at high significance.
The contribution to the ISW effect due to structures at low redshift was investigated as well\cite{francis2010}.  Using a photometric redshift catalogue constructed from 2MASS \citep{2MASS} and SuperCOSMOS  \citep{supercosmos} surveys with a median redshift of $z = 0.08$, an under-density in the galaxy distribution was found in the CS direction. This under density can account for a CMB decrement of $\Delta T \simeq -7 \mu$K in the $\Lambda$-Cold Dark Matter (\LCDM) model, i.e. falling short of fully explaining the CS anomaly. Note that although no void has been found that could fully explain the CS anomaly, there is strong, $\gtrsim 4.4\sigma$, statistical evidence that superstructures imprint on the CMB as cold and hot spots \citep{GranettEtal2008,Planck23,CaiEtal2013}.

The principal scientific question, whether a void causes the CS, can be approached  on three levels: i) is there a low density region (supervoid) in the CS region ii) if there is, how rare is that superstructure iii) is that structure large and deep enough to significantly contribute to the CS in the framework of the (linear) ISW effect in $\Lambda$CDM (or alternative) theories? In particular, if the first question is answered affirmatively, and the resulting void is rare enough that chance alignment with the CS is unlikely, the connection can be established regardless the answer of the third question. Present work is focused on the observational question i) as it can be answered with our data set at high significance. As we show later, we detect a low density region centered on the CS. We also engage question iii) with simultaneous modeling of the underdensity in the galaxy field and the CS pattern observed in CMB maps. On the other hand, we will address ii) at an approximate fashion.

\section{Methods}

Next we describe some of our methods and procedures, while Szapudi etal.\cite{SzapudiEtal2014} and Finelli etal.\cite{FinelliEtal2014} can be consulted for further details.

To probe the low redshift region unconstrained by previous studies, a large area survey is needed reaching up to $z \simeq 0.3$. The redshift distribution of the recently produced WISE-2MASS catalog \citep{KovacsSzapudi2014} has a median redshift of $z\simeq 0.14$, and covers 21,200 square degrees after masking dusty regions. The catalog contains sources to flux limits of $W1_{WISE}$ $\leq $ 15.2 mag and $J_{2MASS}$  $\leq $ 16.5 mag. These conservative limits result in a dataset deeper than 2MASS and more uniform than WISE\cite{wise}.

Galaxies are biased tracers of the underlying dark matter distribution, we thus measured and modeled the angular power spectrum of WISE-2MASS galaxies, and found a bias $b_{g} = 1.41 \pm 0.07$. We look for underdensity at the CS in this projected density field, and scan the full observable sky in order to identify the largest underdensities at $z < 0.3$.

We then set out to create a detailed map of the CS region in three dimensions. Therefore, WISE-2MASS galaxies have been matched with Pan-STARRS1\cite{ps1ref} (PS1) objects within a $50^\circ\times50^\circ$ area centred on the CS, except for a $\rm{Dec}\geq-28.0$ cut to conform to the PS1 boundary. We used PV1.2 reprocessing  of PS1 in an area of 1,300 square degrees.
For PS1, we required a proper measurement of Kron and PSF magnitudes in $g_{\rm P1}$, $r_{\rm P1}$ and $i_{\rm P1}$ bands that  were used to construct photometric redshifts (photo-$z$'s) with a Support Vector Machine algorithm. The training and control sets were created matching WISE-2MASS, PS1, and the Galaxy and Mass Assembly \cite{gama} (GAMA) redshift survey. Our three dimensional catalog of WISE-2MASS-PS1 galaxies contains photo-$z$'s with an estimated error of $\sigma_{z} \approx 0.034$. Redshift distribution of our data set is shown in the upper left panel of Fig.~\ref{fig_pz}.

For our measurements, we take the centre and the size of the CS from the literature as measured in the CMB temperature maps. In particular we use the latest  Planck  results \citep{Planck23} for the centre $(l,b) \simeq (209^\circ,-57^\circ)$ in Galactic coordinates, and  $R=5^\circ$ \citep{VielvaEtal2003, CruzEtal2004} and $R=15^\circ$ \citep{InoueEtal2010,SmithHuterer2010} radii for the size. The $R \leq 5^\circ$ region is the most anomalous part of the CS region, while an outer hot ring at $R \approx 15^\circ$ makes the whole structure more significant \citep{SmithHuterer2010}. 

\begin{figure}
\begin{center}
\includegraphics[width=70mm]{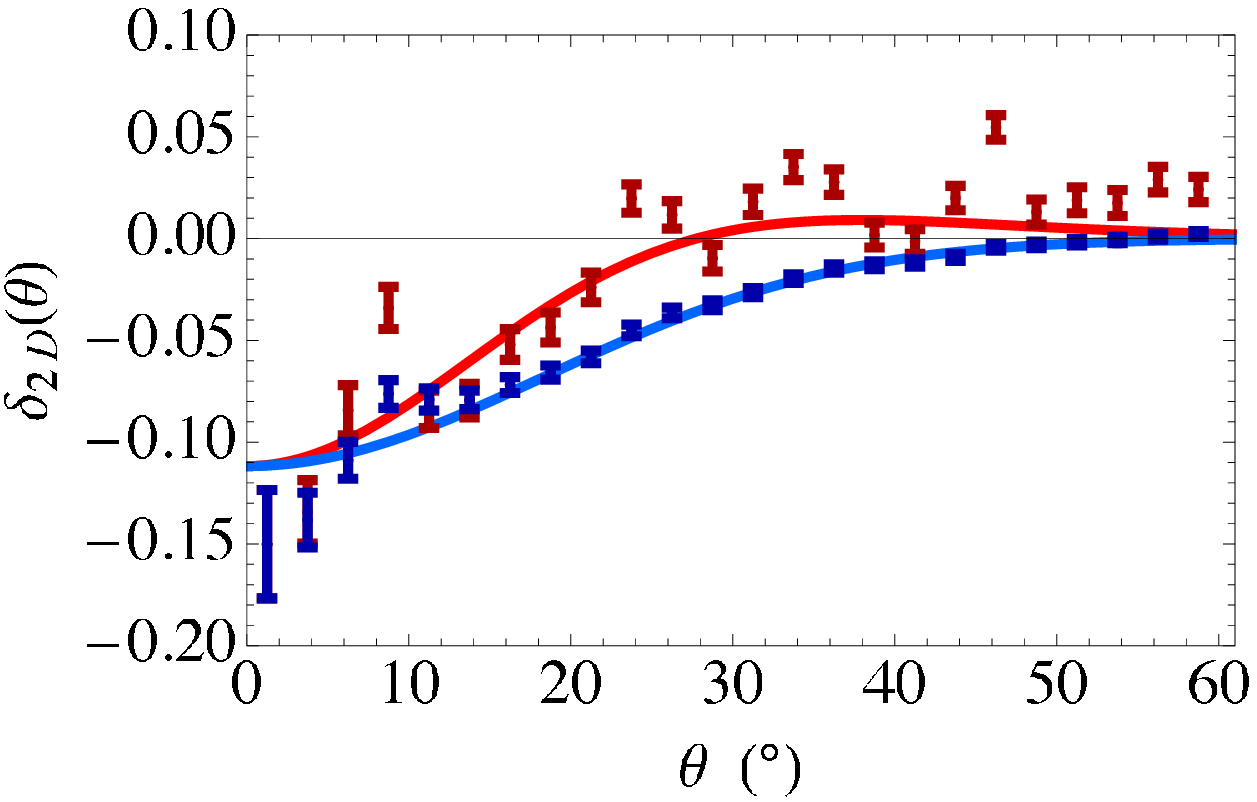}
\includegraphics[width=70mm]{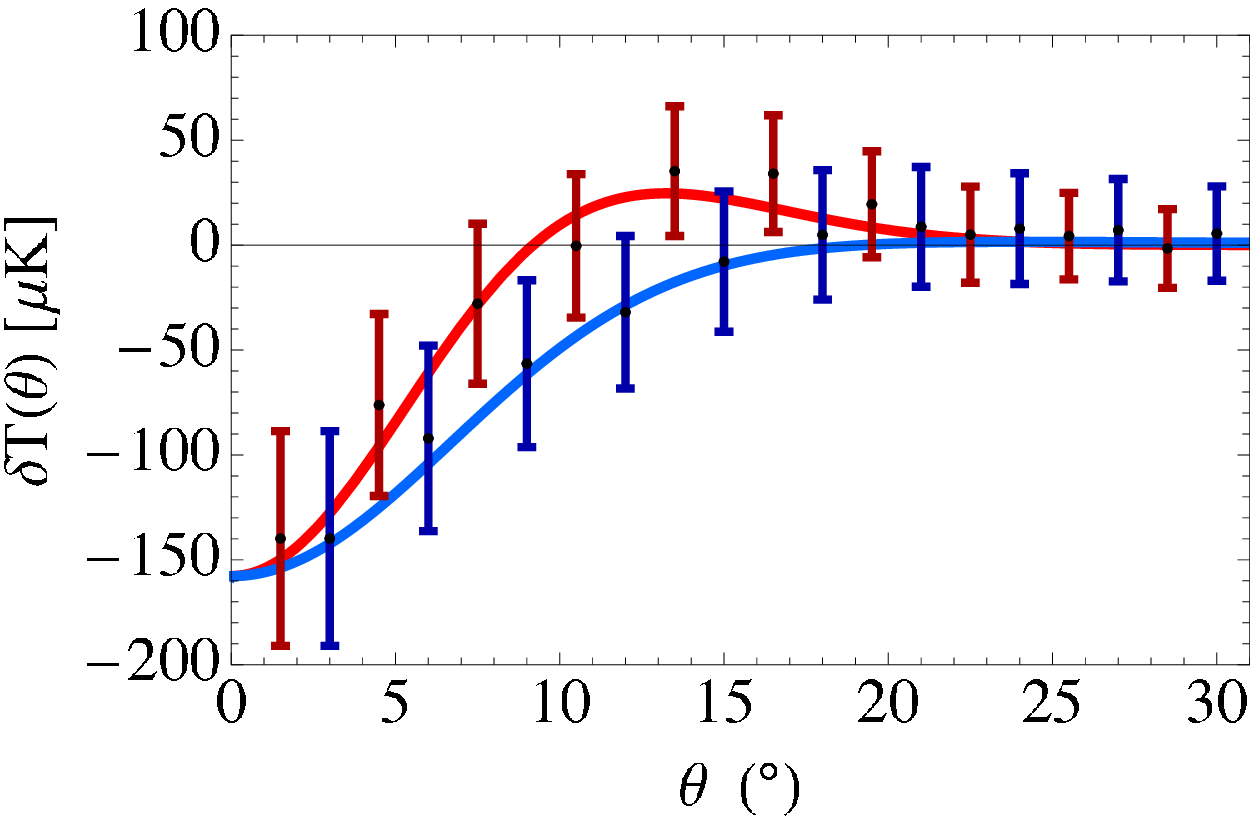}
\caption{The density profile from WISE-2MASS catalogue compared with the theoretical model
for the underdensity (\ref{profile_WISE}) (left panel). The temperature profile from Planck SMICA map (right panel) compared with the 
predicted signal (\ref{profile_CMB}). The red (blue) lines are the theoretical profiles for 
rings (disks) and in dark red (blue) are the measurements.}
\label{fig_prof}
\end{center}
\end{figure}

\section{Measurements with WISE-2MASS}

The projected radial profile centred on the CS is shown in the left panel of Fig.~\ref{fig_prof}. Measurement errors are due to Poisson fluctuations calculated from the expected number of galaxies in a ring or a disk.  We detect a low density region with overwhelming significance. In particular,  at our pre-determined radii,  $5^\circ$ and $15^\circ$,  we find signal-to-noise ratios $S/N \sim 12$ for rings.  The size of the underdense region is surprisingly large: it is detected to  $\sim 20^\circ$ with high ($\gtrsim 5\sigma$) significance. At larger radii, the radial profile is  consistent with a supervoid surrounded by a gentle compensation (overdensity) that converges to the average galaxy density at $\sim 50^\circ$ \cite{PapaiSzapudi2010}. The supervoid might also contain a deeper inner void, with its own compensation at around $\simeq 8^\circ$.

The underdensity in WISE-2MASS is modeled within a $\Lambda$LTB framework \citep{GBH2008}, 
assuming a spatial curvature profile $k(r)=k_0\,r^2\exp(-r^2/r_0^2)$. A linear metric perturbation in $\Lambda$CDM is defined as 
\be
\Phi(\tr) = \Phi_0 \,\tau^2\,e^{-\frac{\tr^2}{\tr_0^2}} \,,
\label{profile_Phi}
\ee
with an LTB radius $r$ defined as $\tr=\sqrt{3/4\pi}\,H_0r$ using the co-moving FRW radius and $\tau$ conformal time. 
A 3D density profile for the void corresponds to this scalar potential, given as
\be
\delta(\tr) = - \delta_0 \left( 1 - \frac{2\tr^2}{3\tr_0^2} \right)\,e^{-\frac{\tr^2}{\tr_0^2}}\,,
\label{profile3D}
\ee
and characterized by two parameters, co-moving width $\tr_0$ and depth $\delta_0$. Then the 3D density profile (\ref{profile3D}) is projected onto the transverse plane using the WISE-2MASS window 
function $\phi(y)$, with the center of the supervoid at co-moving distance $y_0$, and $\tr^2(y,\theta)=y^2+y_0^2-2y y_0\,\cos\theta$,
\be
\delta_{2D}(\theta) = \int_0^\infty \delta \left(\tr(y,\theta)\right)\,y^2\phi(y)dy\,.
\label{profile_WISE}
\ee
\begin{figure}
\begin{center}
\includegraphics[width=70mm]{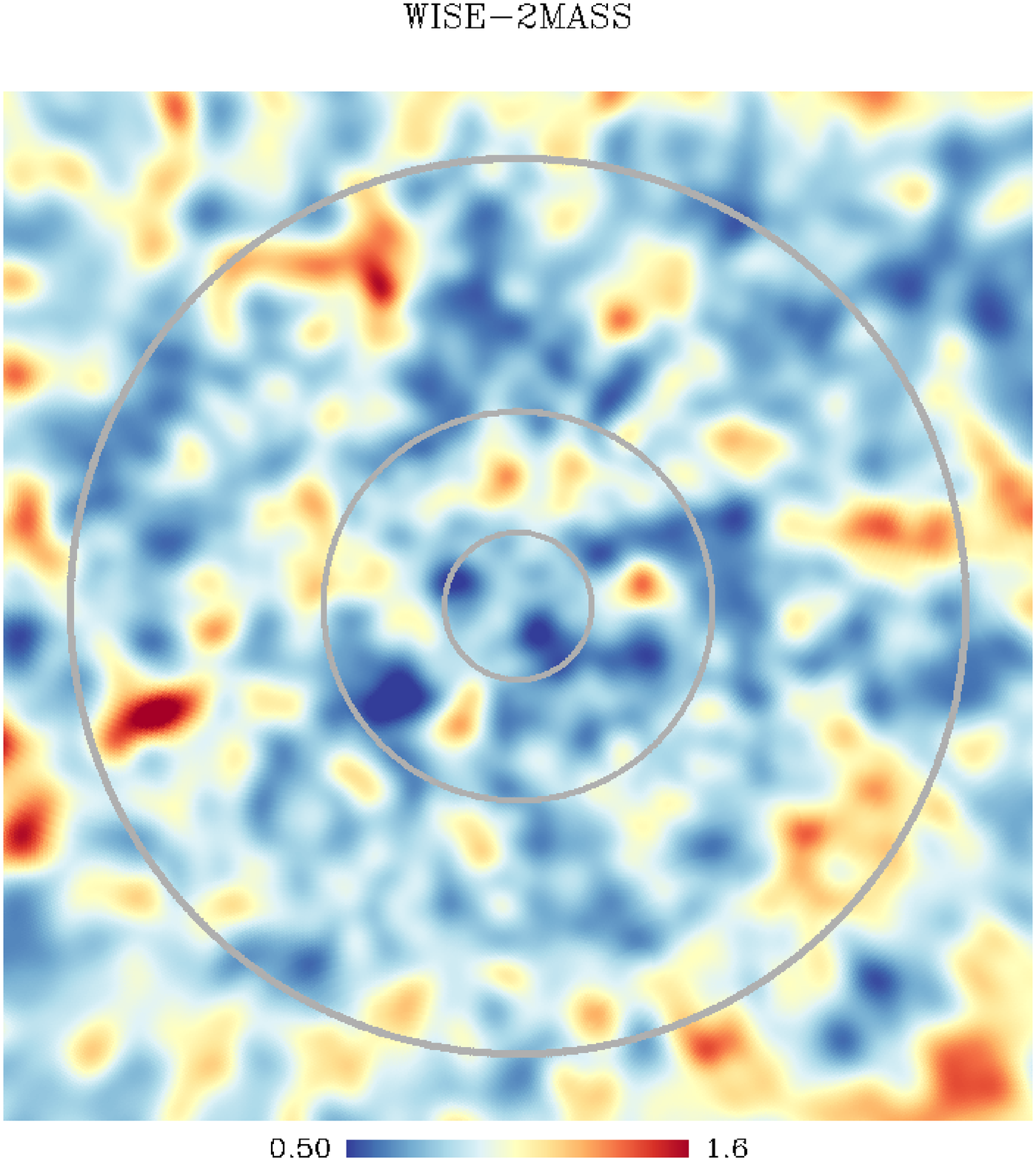}
\includegraphics[width=70mm]{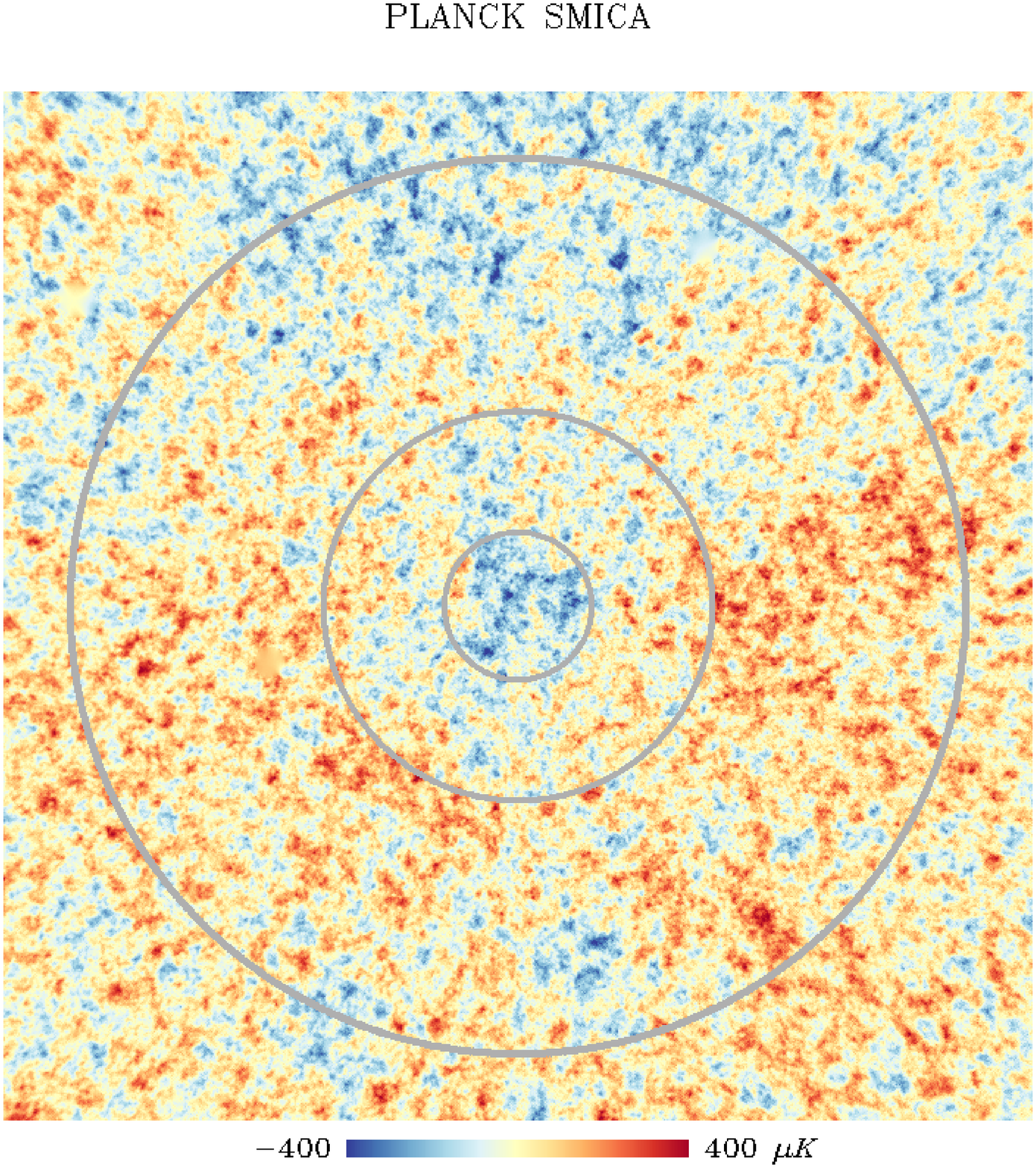}
\caption{The WISE-2MASS (left) and Planck SMICA (right) maps in the direction of the Cold Spot. Circles correspond to 
$5^{\circ}$, $14^{\circ}$ and $29^{\circ}$ radii.}
\label{fig_ltb2}
\end{center}
\end{figure}
The linear ISW and the non-linear RS effects are derived from the metric perturbation (\ref{profile_Phi}). Note that effect on the CMB is dominated by the non-linear Rees-Sciama effect for a large compensated void with a profile of Eq.~\ref{profile3D}, as given by
\be
\delta T (\theta) = - A  \left(1- \frac{28}{13} \frac{\theta^2}{\tilde\theta_0^2} \right)\,e^{-2\frac{\theta^2}{\tilde\theta_0^2}}\,,
\label{profile_CMB}
\ee
where $\tilde\theta_0=\sqrt{3/4\pi}\,\theta_0$, assuming small angle approximation, $\tan\theta\simeq \theta$. The ISW effect is proportional to the time derivative of the potential, thus usually smoother then the density field. The RS effect, however, depends on higher derivatives, thus the size of the CMB pattern is more compact then the size of the void.  The actual coupling is described by a $\sqrt{3/4\pi}\simeq 0.48$ scaling factor between the scales of the underdensity and the corresponding pattern on the CMB.
The magnitude of the CMB temperature depression depends on the parameters of the void,
\be\label{Amp}
A = 51.0 \,\mu{\rm K}\,\left(\frac{r_0}{155.3\,h^{-1}{\rm Mpc}}\right)^3 \left(\frac{\delta_0}{0.2}\right)^2\,,
\ee
and $\theta_0 = (180^\circ/\pi)(r_0/d_A(z_0))$, with $d_A(z)$ the angular diameter distance 
in a flat $\Lambda$CDM model ($\Omega_{\rm M} = 0.3, h=0.7$), and $z_0$ the redshift of the 
center of the void, at co-moving distance $y_0 = y(z_0)$.

A large LTB void can explain a CMB cold spot of about half the size, or the same decrement can be explained by a shallower void than is allowed by $\Lambda$CDM cosmology. Note that the LTB model is only used as a general relativistic model for the void dynamics and its effect on the CMB, while the background cosmology is assumed to be standard $\Lambda$CDM.

For detailed measurements, a $\chi^2$ statistic is constructed. We perform a simultaneous fit for the projected LTB void in the WISE-2MASS map, and the 
corresponding temperature depression effect in the CMB data.  We carry out a 3-parameter fit $\delta_0$, $r_0$, and $z_0$, which are fitted by the following statistic:
\ba
&&\hspace{-1.4cm}\chi^2_{_{\rm tot}}(\delta_0, r_0, z_0) = \chi^2_{_{\rm LSS}} + \chi^2_{_{\rm CMB}}\,, 
\label{chi2tot} \\[1mm]
\chi^2_{_{\rm LSS}}\!&\!=\!&\!\sum_i \frac{\left(\delta_{2D}(\theta_i) - \delta_i^{\rm LSS}\right)^2}{\sigma_i^2} \,,
\label{chi2lss} \\[1mm]
\chi^2_{_{\rm CMB}}\!&\!=\!&\!\sum_{ij} \Big(\delta T(\theta_i) \! - \! \delta T_i^{\rm CMB}\Big)C_{ij}^{-1}
\Big(\delta T(\theta_j) - \delta T_j^{\rm CMB}\Big)\,. \label{chi2cmb}
\ea
The first term corresponds to the $\chi^2$ of the projected LTB void profile (\ref{profile_WISE}) in the WISE-2MASS density field, using uncorrelated Poisson errors, $\sigma_i$.  The second term is the $\chi^2$ of the CMB profile compared with the LTB prediction (\ref{profile_CMB}) of the void observed in WISE-2MASS. 
Note that Poisson fluctuations are uncorrelated, while the covariance matrix of rings in the CMB indicates correlations, and it was determined from 10,000 Gaussian CMB realizations. 
The best fit LTB void parameters we have found are $\delta_0  = 0.25 \pm  0.10~(1\sigma)$, $r_0=(195 \pm 35)\mpc ~(1\sigma)$, and $z_0 = 0.16 \pm  0.04~(1\sigma)$.

The LTB model parameter $\delta_0$ is the 3D dark matter density, giving a 12\% projected underdensity at the center of the void,  i.e. $\delta_{\rm 2D}(\theta=0) = -0.12$. The angular sizes $\theta_0 = 28.8^\circ\pm 5.2^\circ$, and $\tilde \theta_0 = 14.1^\circ\pm 2.5^\circ$ are derived parameters, which correspond to angular scales of the profile on the galaxy map and the CMB, respectively. Note that the radius of the LTB profile $\tilde \theta_0 = 14.1^\circ\pm 2.5^\circ$ on the CMB matches the outer hot ring around the cold spot discussed in \cite{ZhangHuterer2010}. For later comparison, we calculate the averaged underdensity within the best fit radius $r_0=195\,h^{-1}$Mpc. The 3D top-hat-averaged density from the LTB profile, see Eq.~(\ref{profile3D}), is $\bar\delta = 3/r_0^3  \int_0^{r_0}  r^2 dr \delta(r) = - \delta_0/e$. This finally gives the average void depth $\bar\delta = - 0.10 \pm 0.03$.

\section{Measurements with WISE-2MASS-Pan-STARRS1}
\begin{figure}
\begin{center}
\includegraphics[height=100mm]{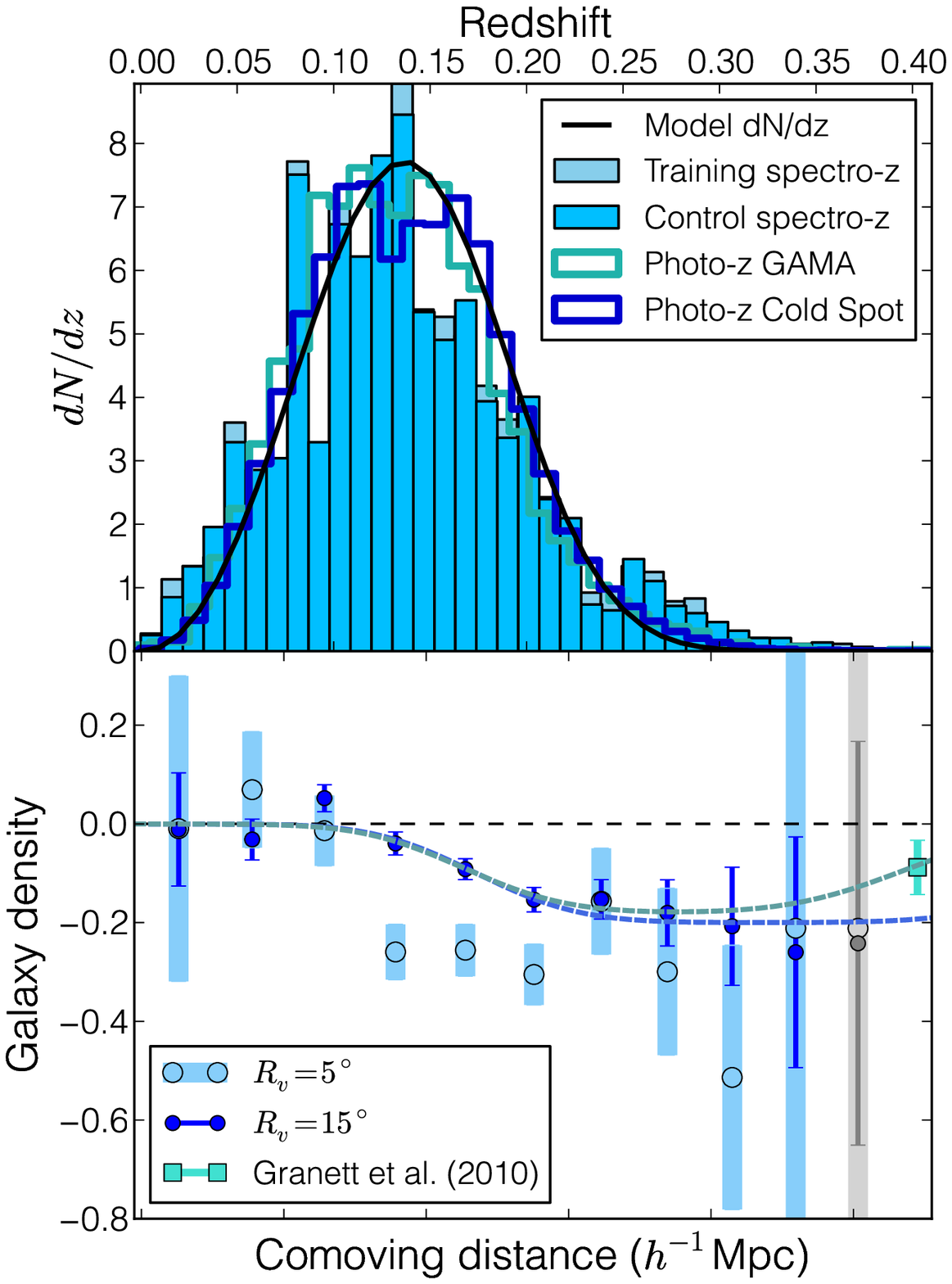}
\includegraphics[height=100mm]{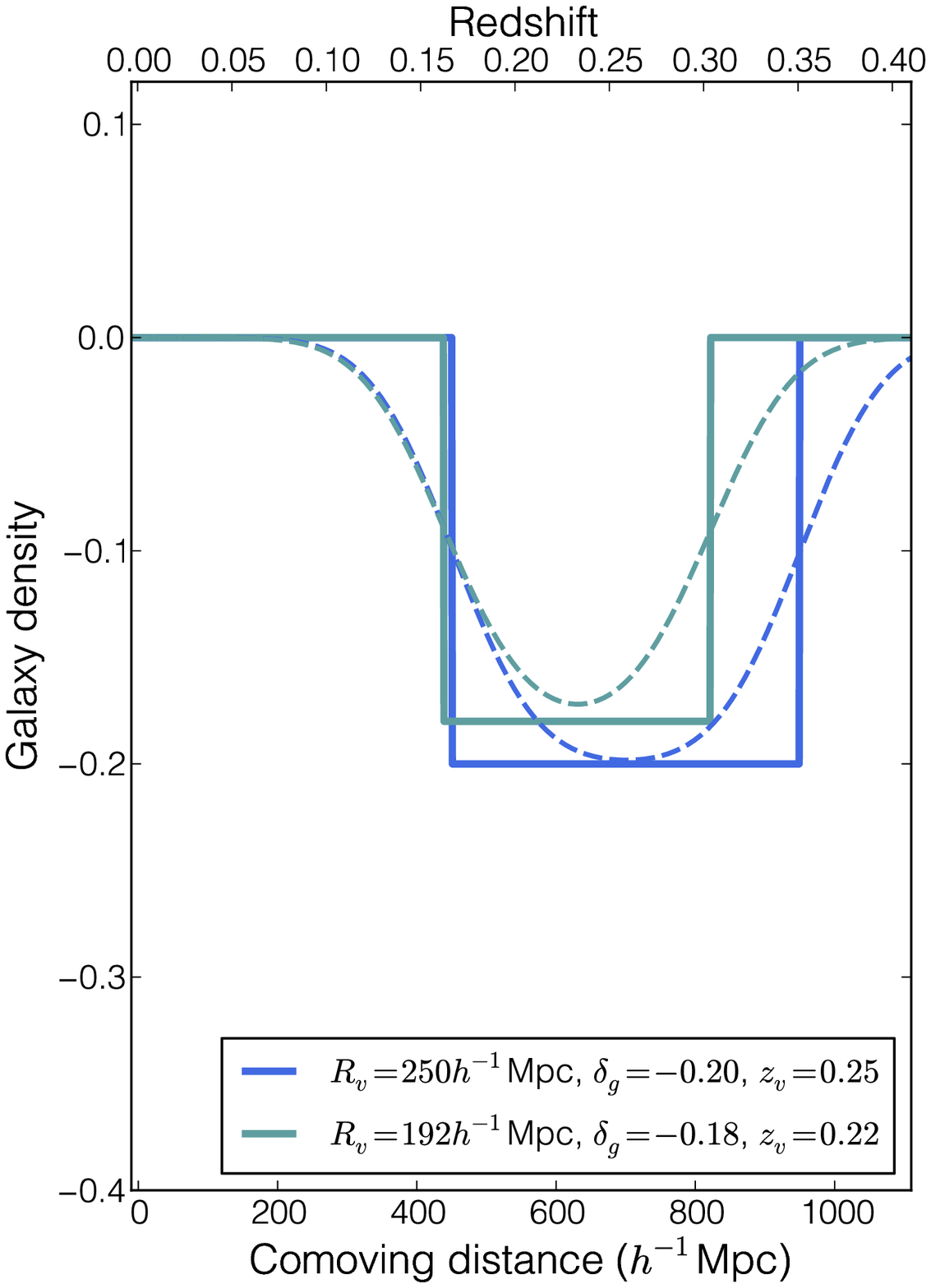}
\caption{Left: Redshift Space Distribution in the Cold Spot Region. The upper panel illustrates the normalized redshift distribution of our subsamples used in the photo-$z$ pipeline; training and control sets selected in GAMA, photo-$z$ distributions estimated for the WISE-2MASS-PS1-GAMA control sample, and photo-$z$'s of interest in the WISE-2MASS-PS1 matched area. The median redshift of all samples is $z\simeq 0.14$. The lower left panel shows galaxy densities measured at the two fiducial radii in the CS region. The blue dashed lines show our best fit top hat toy model that folds in the photometric redshift errors. The green dashed lines illustrate that without the earlier measurement (the rightmost point\cite{GranettEtal2010}) we could not distinguish a much larger void. Right: Top-hat void models. The plot shows the best fit, and a slightly larger 3-parameter top hat toy model for a supervoid. The latter could not be excluded by PS1 data alone.  Dashed curves represent the smearing of the profile by photo-z errors. The model redshift distribution of our observations was multiplied by this smeared profile, and the resulting profile was convolved with the photo-$z$ error.}
\label{fig_pz}
\end{center}
\end{figure}
Photometric redshifts in the WISE-2MASS-PS1 galaxy catalog provide a three-dimensional view of the superstructure.  We count galaxies as a function of redshift in disks centred on the CS using the two pre-determined angular radii, $R=5^\circ$, and $15^\circ$.  The galaxy density calculated from the average redshift distribution is shown in the upper panel of Fig.~\ref{fig_pz}. Note that the larger disk is not fully contained in our photo-$z$ catalog due to the limited PS1 footprint.
In photo-$z$ bins of width of $\Delta z = 0.035$, we found $S/N \sim 5$ and $S/N \sim 6$ for the deepest under-density bins, respectively. The measurement errors are still due to Poisson fluctuations in the redshift bins. To extend our analysis to higher redshifts, we add a previous measurement\cite{GranettEtal2010} in a photo-$z$ bin centred at $z=0.4$.

For a rudimentary understanding of our counts, we build toy models from top-hat voids in the $z$ direction convolved with the photo-$z$ errors. This model has  three parameters, redshift ($z_{\rm void}$), radius ($R_{\rm void}$), and depth ($\delta_{g}$).
We carry out the $\chi^{2}$-based maximum likelihood parameter estimation for $R=15^{\circ}$, using our first 10 bins combined with the extra bin measured by \cite{GranettEtal2010} for $n=11$ data points.   We find $\chi^{2}_{0} = 92.4$ for the null hypothesis of no void. The marginalized results for the parameters of our toy model are $z_{\rm void}=0.22\pm 0.01$ $(2\sigma)$,  $R_{\rm void}=(192 \pm 15)\mpc $ $(2\sigma)$, and $\delta_{g}=-0.18\pm 0.01$ $(2\sigma)$ finding $\chi^{2}_{min} = 5.97$. For $k=3$ parameters, there is $\nu = n-k=8$ degrees of freedom, i.e. the toy model is a surprisingly good fit and clearly preferred over the null hypothesis. We take the galaxy bias into account finding an average depth of $\delta = \delta_{g} / b_{g} \simeq -0.13 \pm 0.03$ ($2\sigma$).

\begin{figure}
\begin{center}
\includegraphics[width=175mm]{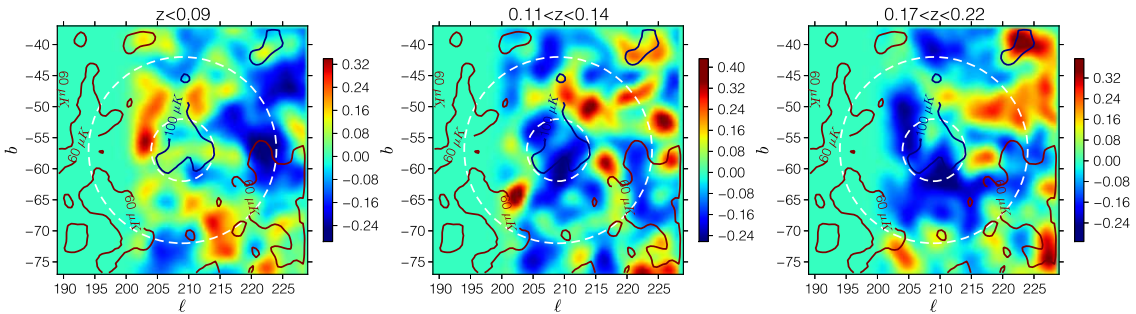}
\caption{Tomographic view of the CS region. Below $z < 0.09$ appears to show the front compensation (higher density area) of the large void, although there is an under dense structure to the right. The slice at $0.11 < z < 0.14$  cuts into front of the supervoid, and the ring round represents a slice of the compensation. Finally the slice $0.17 < z < 0.22$ cuts into the front half of the supervoid with compensation around it. The structure on the upper right reaching into the supervoid might be real, or a shadow of the same structure in the middle figure due to the photometric redshift errors. Note that the left side of the images reflects the mask of the PS1 data set.}
\label{fig_visu}
\end{center}
\end{figure}

While the toy model quantifies the properties of the void, careful inspection of the Figures reveals a complex structure. There appears to be a compensation in front of the supervoid at around $z\simeq 0.05-0.08$, and the significantly deeper counts at the smaller radii show that the void is deeper at the centre. An approximate tomographic imaging of the CS region is presented in Fig.~\ref{fig_visu}. in three slices of $z < 0.09$, $0.11 < z < 0.14$,  $0.17 < z < 0.22$ and smoothed at $2^{\circ}$ scales. \corr{The gap between the slices corresponds to $1\sigma$ photometric redshift errors, and above $z\simeq 0.22$, the middle of the supervoid, the Poisson errors become significant.}{} The Planck  \cite{Planck23} SMICA CMB temperature map is over-plot with contours. While we characterized the profile in radial bins, the geometry of the supervoid is more complex. It is noteworthy that deepest part of the void is close to the centre of the CS in the middle slice. Given the enormous size, the rich structure is expected: the supervoid contains voids, their compensations, filaments, and appears to have its own compensation.

\section{Discussion \& Conclusions}

We have found evidence for a supervoid at redshift $z = 0.22\pm 0.01$ with radius $R = (192\pm 15)\,h^{-1}$Mpc and depth of $\delta = -0.13 \pm 0.03$. These parameters are in excellent agreement with the findings of our WISE-2MASS projected density analysis, with the redshift slightly closer, but only by $1.5\sigma$.  Note that the orthogonal nature of the two measurements suggest spherical symmetry for the void.

We conclude that a non-linear LTB model based on the projected profile in the WISE-2MASS catalog matches well the profile observed on the CMB. We also estimated the linear ISW effect due to such an underdensity from a simple approximation \cite{RudnickEtal2007}. We found that it could significantly affect the CMB,  of order $-20$ $\mu$K, falling short of a full explanation of the CS anomaly. The above  findings address questions i) and iii) in the introduction.

The tomographic maps produced by photometric redshifts from PS1, while not extending to large enough radii at the moment, have an intriguing match with the Cold Spot region as observed on the CMB.  The observed morphological similarity to a compensated surrounding over dense shell, which plausibly would have fragmented into galaxy clusters visible in the projected slices as several "hot spots" surrounding the CS, is noteworthy. 
Based on our toy model, and assuming a simple Gaussian model in \LCDM\ cosmology, we now attempt to answer question ii), i.e the rarity of the supervoid.
We estimate that the supervoid we detected corresponds to a rare, at least $\gtrsim 3.5\sigma$, density fluctuation; thus chance alignment with another rare structure, the CS, is negligible. As a further test, we smoothed the projected WISE-2MASS map with a $25^\circ$ Gaussian finding only one underdensity as significant as the one we discovered in the CS region. This second void, to be followed up in future research, is clearly visible in shallow 2MASS maps \cite{RassatEtal2013,francis2010} as a large underdensity, and in the corresponding reconstructed ISW map \cite{RassatStarck2013} as a cold imprint. Therefore we conclude that it is likely to be closer thus smaller in physical size, and not large enough to leave colder imprints on the CMB.

Any tension with \LCDM, e.g. in the possible rarity of the observed supervoid, could be addressed in models of modified gravity, ordinarily screened in clusters, but resulting in an enhanced growth rate of voids as well as an additional contribution to the ISW signal.

\section*{Acknowledgments}

The authors thank Fabio Finelli and Francesco Paci who had crucial roles in developing the LTB explanation for the Cold Spot\cite{FinelliEtal2014}.
We acknowledge the support of NASA grants NNX12AF83G and NNX10AD53G. AK and ZF acknowledge support from OTKA through grant no. 101666. In addition, AK acknowledges support from the Campus Hungary fellowship program. We make use of HEALPix \citep{healpix} software in our project. 
The Pan-STARRS1 Surveys (PS1) have been made possible through contributions by the Institute for Astronomy, the University of Hawaii, the Pan-STARRS Project Office, the Max-Planck Society and its participating institutes, the Max Planck Institute for Astronomy, Heidelberg and the Max Planck Institute for Extraterrestrial Physics, Garching, The Johns Hopkins University, Durham University, the University of Edinburgh, the Queen's University Belfast, the Harvard-Smithsonian Center for Astrophysics, the Las Cumbres Observatory Global Telescope Network Incorporated, the National Central University of Taiwan, the Space Telescope Science Institute, and the National Aeronautics and Space Administration under Grant No. NNX08AR22G issued through the Planetary Science Division of the NASA Science Mission Directorate, the National Science Foundation Grant No. AST-1238877, and the University of Maryland, and Eotvos Lorand University (ELTE).


\begin{thebibliography}{}

\end{thebibliography}


\begin{thebibliography}{10}

\bibitem{bennett2012}
C.~L. {Bennett}, D.~{Larson}, and J.~L et~al. {Weiland}.
\newblock {\em ArXiv e-prints}, December 2012.

\bibitem{VielvaEtal2003}
P.~{Vielva} et al.
\newblock {\em \apj}, 609:22--34, July 2004.

\bibitem{CruzEtal2004}
M.~{Cruz} et al.
\newblock {\em \mnras}, 356:29--40, January 2005.

\bibitem{Planck23}
{Planck 2013 results. XXIII.}
\newblock {\em ArXiv e-prints}, March 2013.

\bibitem{CruzEtal2008}
M.~{Cruz} et al.
\newblock {\em \mnras}, 390:913--919, November 2008.

\bibitem{Vielva2010}
P.~{Vielva}.
\newblock {\em Advances in Astronomy}, 2010, 2010.

\bibitem{InoueSilk2006}
K.~T. {Inoue} and J.~{Silk}.
\newblock {\em \apj}, 648:23--30, September 2006.

\bibitem{InoueSilk2007}
K.~T. {Inoue} and J.~{Silk}.
\newblock {\em \apj}, 664:650--659, August 2007.

\bibitem{InoueEtal2010}
K.~T. {Inoue}, N.~{Sakai}, and K.~{Tomita}.
\newblock {\em \apj}, 724:12--25, November 2010.

\bibitem{Inoue2012}
K.~T. {Inoue}.
\newblock {\em \mnras}, 421:2731--2736, April 2012.

\bibitem{SachsWolfe}
R.~K. {Sachs} and A.~M. {Wolfe}.
\newblock {\em ApJL}, 147:73, January 1967.

\bibitem{ReesSciama}
M.~J. {Rees} and D.~W. {Sciama}.
\newblock {\em \nat}, 217:511--516, February 1968.

\bibitem{RudnickEtal2007}
L.~{Rudnick}, S.~{Brown}, and L.~R. {Williams}.
\newblock {\em \apj}, 671:40--44, December 2007.

\bibitem{SmithHuterer2010}
K.~M. {Smith} and D.~{Huterer}.
\newblock {\em \mnras}, 403:2--8, March 2010.

\bibitem{BremerEtal2010}
M.~N. {Bremer} et al.
\newblock {\em \mnras}, 404:L69--L73, May 2010.

\bibitem{GranettEtal2010}
B.~R. {Granett}, I.~{Szapudi}, and M.~C. {Neyrinck}.
\newblock {\em \apj}, 714:825--833, May 2010.

\bibitem{PapaiSzapudi2010}
P. {P{\'a}pai}, and I. {Szapudi}.
\newblock {\em \apj}, 725:2078--2086, 2010


\bibitem{francis2010}
C.~L. {Francis} and J.~A. {Peacock}.
\newblock {\em \mnras}, 406:14--21, July 2010.

\bibitem{SzapudiEtal2014}
Szapudi, I. and Kov{\'a}cs, A., Granett, B.~R., et al.
\newblock  arXiv:1405.1566, 2014.

\bibitem{FinelliEtal2014}
Finelli, F.,  Garcia-Bellido, J., Kovacs, A., Paci, F., and Szapudi, I.
\newblock arXiv:1405.1555, 2014.

\bibitem{2MASS}
M.~F. {Skrutskie}, R.~M. {Cutri}, R.~{Stiening}, and et~al.
\newblock {\em \aj}, 131:1163--1183, February 2006.

\bibitem{supercosmos}
N.~C. {Hambly} and et~al.
\newblock {\em \mnras}, 326:1279--1294, October 2001.

\bibitem{GranettEtal2008}
B.~R. {Granett}, M.~C. {Neyrinck}, and I.~{Szapudi}.
\newblock {\em \apjl}, 683:L99--L102, August 2008.

\bibitem{CaiEtal2013}
Y.-C. {Cai} et al.
\newblock {\em ArXiv e-prints}, January 2013.

\bibitem{KovacsSzapudi2014}
A.~{Kov{\'a}cs} and I.~{Szapudi}.
\newblock {\em ArXiv e-prints}, December 2014.

\bibitem{wise}
E.~L. {Wright} et~al.
\newblock {\em \aj}, 140:1868--1881, December 2010.

\bibitem{ps1ref}
N.~{Kaiser}.
\newblock In {\em SPIE Conference Series}, 2004.

\bibitem{gama}
S.~P. {Driver}, D.~T. {Hill}, and et~al.
\newblock {\em \mnras}, 413:971--995, May 2011.

\bibitem{GBH2008}
J.~{Garcia-Bellido} and T.~{Haugb{\o}lle}.
\newblock {\em JCAP}, 4:3, April 2008.

\bibitem{ZhangHuterer2010}
R.~{Zhang} and D.~{Huterer}.
\newblock {\em Astroparticle Physics}, 33:69--74, March 2010.

\bibitem{RassatEtal2013}
A.~{Rassat}, J.-L. {Starck}, and F.-X. {Dup{\'e}}.
\newblock {\em \aap}, 557:A32, September 2013.

\bibitem{RassatStarck2013}
A.~{Rassat} and J.-L. {Starck}.
\newblock {\em \aap}, 557:L1, September 2013.

\bibitem{DES}
{The Dark Energy Survey Collaboration}.
\newblock {\em ArXiv Astrophysics e-prints}, October 2005.

\bibitem{healpix}
K.~M. {Gorski}, E.~{Hivon}, and et~al.
\newblock {\em \apj}, 622:759--771, April 2005.
\end{thebibliography}
\end{document}